\documentclass[a4paper,fleqn]{cas-sc}

\usepackage[authoryear,longnamesfirst]{natbib}
\usepackage{algorithm}
\usepackage[utf8]{inputenc}
\usepackage[noend]{algpseudocode}

\def\tsc#1{\csdef{#1}{\textsc{\lowercase{#1}}\xspace}}
\tsc{WGM}
\tsc{QE}
\tsc{EP}
\tsc{PMS}
\tsc{BEC}
\tsc{DE}

\begin{document}
\let\WriteBookmarks\relax
\def\floatpagepagefraction{1}
\def\textpagefraction{.001}

\shorttitle{Trusted-On-Demand-FL}

\shortauthors{Mario Chahoud et~al.}

\title [mode = title]{Trust Driven On-Demand Scheme for Client Deployment in Federated Learning}                      




%
\author[1,2,3]{Mario Chahoud}[orcid=0000-0002-6070-3133]



\ead{mario.chahoud@concordia.ca}


\credit{Conceptualization; Data curation; Formal analysis; Funding acquisition; Investigation; Methodology; Project administration; Resources; Software; Supervision; Validation; Visualization; Roles/Writing - original draft; Writing - review \& editing}

\affiliation[1]{Artificial Intelligence & Cyber Systems Research Center, Department of CSM, Lebanese American University, Beirut, Lebanon}
\affiliation[2]{Mohammad Bin Zayed University of Artificial Intelligence, Abu Dhabi, UAE}

\affiliation[3]{Concordia Institute for Information Systems Engineering, Concordia University, Montreal, Canada}

\affiliation[4]{KU 6G Research Center, Department of CS, Khalifa University, UAE}

\affiliation[5]{Center of Cyber-Physical Systems (C2PS), Department of CS, Khalifa University, Abu Dhabi, UAE}

\author%
[4,1]
{ Azzam Mourad}[orcid=0000-0001-9434-5322]
\ead{azzam.mourad@lau.edu.lb}
\credit{Conceptualization; Formal analysis; Funding acquisition; Investigation; Methodology; Project administration; Resources; Supervision; Roles/Writing - original draft; Writing - review \& editing}

\author%
[5]
{ Hadi Otrok}[orcid=0000-0002-9574-5384]
\ead{hadi.otrok@ku.ac.ae}
\credit{Conceptualization; Formal analysis; Funding acquisition; Investigation; Methodology; Project administration; Resources; Supervision; Roles/Writing - original draft; Writing - review \& editing}

\author[4,3]{ Jamal Bentahar}[orcid=0000-0002-3136-4849]

\ead{jamal.bentahar@concordia.ca}

\credit{Conceptualization; Formal analysis; Funding acquisition; Investigation; Methodology; Project administration; Resources; Supervision; Roles/Writing - original draft; Writing - review \& editing}

\author[2]{ Mohsen Guizani}[orcid=0000-0002-8972-8094]

\ead{mguizani@ieee.org}

\credit{Conceptualization; Formal analysis; Funding acquisition; Investigation; Methodology; Project administration; Resources; Supervision; Roles/Writing - original draft; Writing - review \& editing}

\begin{abstract}
Containerization technology plays a crucial role in Federated Learning (FL) setups, expanding the pool of potential clients and ensuring the availability of specific subsets for each learning iteration. However, doubts arise about the trustworthiness of devices deployed as clients in FL scenarios, especially when container deployment processes are involved. Addressing these challenges is important, particularly in managing potentially malicious clients capable of disrupting the learning process or compromising the entire model.
In our research, we are motivated to integrate a trust element into the client selection and model deployment processes within our system architecture. This is a feature lacking in the initial client selection and deployment mechanism of the On-Demand architecture. We introduce a trust mechanism, named "Trusted-On-Demand-FL", which establishes a relationship of trust between the server and the pool of eligible clients. Utilizing Docker in our deployment strategy enables us to monitor and validate participant actions effectively, ensuring strict adherence to agreed-upon protocols while strengthening defenses against unauthorized data access or tampering.
Our simulations rely on a continuous user behavior dataset, deploying an optimization model powered by a genetic algorithm to efficiently select clients for participation. By assigning trust values to individual clients and dynamically adjusting these values, combined with penalizing malicious clients through decreased trust scores, our proposed framework identifies and isolates harmful clients. This approach not only reduces disruptions to regular rounds but also minimizes instances of round dismissal, Consequently enhancing both system stability and security.
\end{abstract}


\begin{highlights}
\item Introducing a Trusted Mechanism for the On-Demand Federated Learning architecture.
\item Recruiting untrusted devices as clients in Federated Learning imposes trust checks.
\item Implementing a dynamic trust evaluation system to continuously assess Trust values.
\item Leveraging bootstrapping and heuristics techniques for client selection.
\item Notably increasing the pool of trusted clients available for Federated Learning.
\end{highlights}


\maketitle

\section{Introduction}
Businesses and software solutions are in a high state of evolution, continually harnessing technology to enhance user experiences. Artificial Intelligence (AI) has emerged as a central player in this progression, revolutionizing various domains. Notably, Federated Learning (FL) has garnered attention as a promising method for implementing AI. FL empowers Machine Learning (ML) models to undergo training on decentralized datasets, eliminating the need for centralizing data sources. This approach not only enhances efficiency but also prioritizes privacy, a critical concern in today's data-driven landscape. The process begins by selecting a subset of available devices to participate in iterative learning stages, ensuring the model's convergence. Subsequently, clients share their updated weights with a central server, which aggregates these weights to refine the model. The selection of clients for each round employs a random selection approach \cite{mna1}.

The act of sharing private data due to privacy concerns poses a significant challenge. However, FL offers a solution by enabling the utilization of newly generated data while preserving confidentiality. Expanding on this concept, recent studies, such as \cite{hani1}, have leveraged containerization technology to efficiently distribute services across nearly-deployed fog devices. Building upon this foundation, subsequent research presented in \cite{mario1kk} and \cite{firstpaperr} extended the architecture to facilitate On-Demand client deployment, allowing for the utilization of private data in the learning process as needed. This On-Demand architecture relies on deploying Docker containers on volunteering devices, which serve as clients in FL rounds. 
This approach has proven effective, particularly in dynamic environments or scenarios with limited available clients, as we now have a larger pool of clients to choose from and more data available for our learning process.
Docker, a leading containerization solution, simplifies application deployment across diverse environments by packaging applications and their dependencies into self-contained, lightweight executable units known as containers. Complementing Docker, Kubeadm offers a utility for swiftly constructing Kubernetes clusters, enabling the creation of clusters using any resource-constrained device and a master node. This flexibility makes Kubeadm a valuable tool for deploying and managing Kubernetes clusters efficiently.

Deploying models encapsulated in our containers onto unknown devices to prepare them to serve as clients raises concerns about the trustworthiness of these devices \cite{basedconference}. Within our FL architecture, this introduces a potential vulnerability in the form of malicious client activities, capable of disrupting the training process.
Leveraging Docker containers proves to be a valuable strategy in addressing trust-related challenges. By encapsulating the model and its dependencies within a Docker container, FL aaplications can ensure the reproducibility and verification of the training process \cite{mariofinals}. Additionally, Docker's isolation mechanisms enhance the security of the training environment, mitigating potential attacks or malicious behaviors from participants. However, caution must be exercised when deploying containers on devices involved in FL, considering the potential for malicious intent \cite{wazz}.
Henthfore, it is essential to introduce trust-oriented mechanisms into both the FL and containerization processes, an aspect often overlooked in prior research endeavors such as those outlined in \cite{lit1}, \cite{lit2}, \cite{newtointro1}, and \cite{lit5}, which primarily focused on addressing issues related to model updates.

The existing body of research offers several methodologies for identifying malicious clients, which tackle many of the challenges mentioned earlier. However, none of the previous studies have put forth the idea of including a trust metric when deploying models On-Demand via containerization technology. 
Within this paper, we illustrate how this trust factor undergoes continuous updates, drawing from the client's successful deployment history. This update process involves a two-step verification mechanism designed to detect any label manipulation or unexpected updates, monitoring for substantial deviations from expected group behaviors in instances of random or consistent weight adjustments, and cross-referencing contextual information to identify any inconsistencies. 
Moreover, it captures the developing nature of trust connections and makes more accurate and informed decisions in the FL process by including dynamic trust calculations while assigning initial trust values using Bootstrapping \cite{bosstr}. For the purpose of deploying models and efficiently selecting clients, we establish a multi-objective formulation that encompasses various objectives and limitations to be taken into account during the deployment process. Solving this multi-objective problem of efficient deployment is achieved through the utilization of a genetic algorithm.
To demonstrate the value and benefits of our method, we employ the MDC dataset \cite{mario14} as an example of real-world simulation, while also utilizing ModularFed \cite{arafehlocalfedd}  to facilitate the application of FL models. 

In this study, various scenarios were examined, including those involving FL setups with static, distant, and unavailable trusted clients. In this context, our dynamic deployment approach, while mindful of our objectives and constraints, empowers capable and dependable clients to actively participate in the learning environment.
Furthermore, several scenarios were conducted to detect the malicious clients and their effect on the model while using the proposed model compared to other solutions in the literature.
The experiments performed produced promising outcomes, showcasing a reduction in the number of rounds needed for convergence, maximizing accuracy, and the successful identification of malicious clients attempting to compromise the model when contrasted with the conventional random selection approach. In summary, this paper's key contributions can be outlined as follows:

\begin{enumerate}
     \item Utilizing a trust-based architecture, trusted clients can be deployed On-Demand in pre-configured FL areas, ensuring secure and reliable operations while leveraging containerization technology .
    \item Employing a continuous monitoring approach, real-time interactions, behavior, and contributions are observed to generate trust values for each client, while efficiently assigning initial values. 
    \item We incorporate a Genetic Algorithm (GA) to enhance the deployment procedure, tackling it as a multi-objective optimization problem and leveraging the evolutionary tactics inherent to the GA.
    
\end{enumerate}

The rest of this article is organized as follows: In Section II, an overview of the current literature is presented. Section III details the architecture and approach of the proposed client deployment strategy. Our dependable formulation for On-Demand-FL is described in Section IV. The genetic algorithm is detailed in Section V. The experiments, results, and evaluation are demonstrated in Section VI, followed by the conclusion in Section VII.

\section{Literature review}

Researchers have conducted various innovative strategies to protect FL applications from malicious attacks. For instance, FLDetector \cite{lit1} scrutinizes model updates to identify malevolent clients, while \cite{lit2} employs a conditional variational autoencoder to detect suspicious updates. Extracting key features and analyzing their similarity using Hamming distance is explored in \cite{lit3}, and \cite{lit4} evaluates received gradients for relevance and similarity data.
Addressing poisoning attacks, \cite{lit5} introduces a resilient federated averaging method, employing a game-based approach where the server and clients participate to detect spurious updates. On the contrary, FedInv \cite{lit6} tackles Byzantine attacks by reversing local updates, creating dummy datasets. Trust-driven client selection is emphasized in \cite{lit7}, focusing on IoT devices, while \cite{lit8} addresses the cold-start problem in FL by leveraging trust.
Further, \cite{paper5} proposes a classification framework combining FL with secure semi-supervised ensemble learning to detect malicious applications while preserving user privacy. Decentralized FL methods, such as the one presented in \cite{paper16}, operate without a central server, relying on online Push-Sum algorithms to establish trust among clients.
In addition, \cite{lit9} suggests a server-free decentralized FL method based on one-way trust relationships among clients, akin to a social network. Lastly, \cite{lit10} introduces the Adaptive Federated Averaging strategy, resilient to Byzantine attacks, utilizing Hidden Markov Models to detect and discard erroneous updates.

While previous research has made significant strides in fortifying FL against various malicious threats, a critical yet unexplored aspect is the integration of trust metrics into client selection and deployment within the On-demand FL architecture. This entails evaluating potential clients' trustworthiness against predefined criteria prior to their selection in FL rounds, thus establishing a pool of newly trusted clients to enhance the overall security and reliability of the FL system. While prior studies predominantly focused on modified models and updates without explicitly addressing the challenge of distinguishing between malicious and benign participants, our approach tailors trust levels to each participant based on their individual characteristics and historical performance. Notably, in \cite{lit7}, the client selection method used incorporated a trust factor. However, such existing investigations have not sufficiently addressed the adaptability of FL models to accommodate newly deployed trusted devices, particularly in environments where such devices are limited, thereby influencing learning dynamics. Moreover, none of the prior works have provided a means to rapidly deploy clients in real-time while considering their behavioral patterns and trustworthiness. 
By integrating trust values into the client selection and deployment mechanisms of our proposed On-Demand architecture, we not only mitigate the risks associated with malicious activities but also cultivate a culture of collaboration and cooperation among participants. This not only strengthens the resilience of FL systems against adversarial attacks but also creates opportunities to generate and deploy new trusted clients for the learning process, thus enhancing the overall effectiveness and adaptability of the FL framework.

\section{Proposed Architecture}
In this section, we present our architecture, followed by descriptions of the functionalities in each component of our architecture.

\begin{figure*}[]
	\centering
	\includegraphics[width=0.8\textwidth]{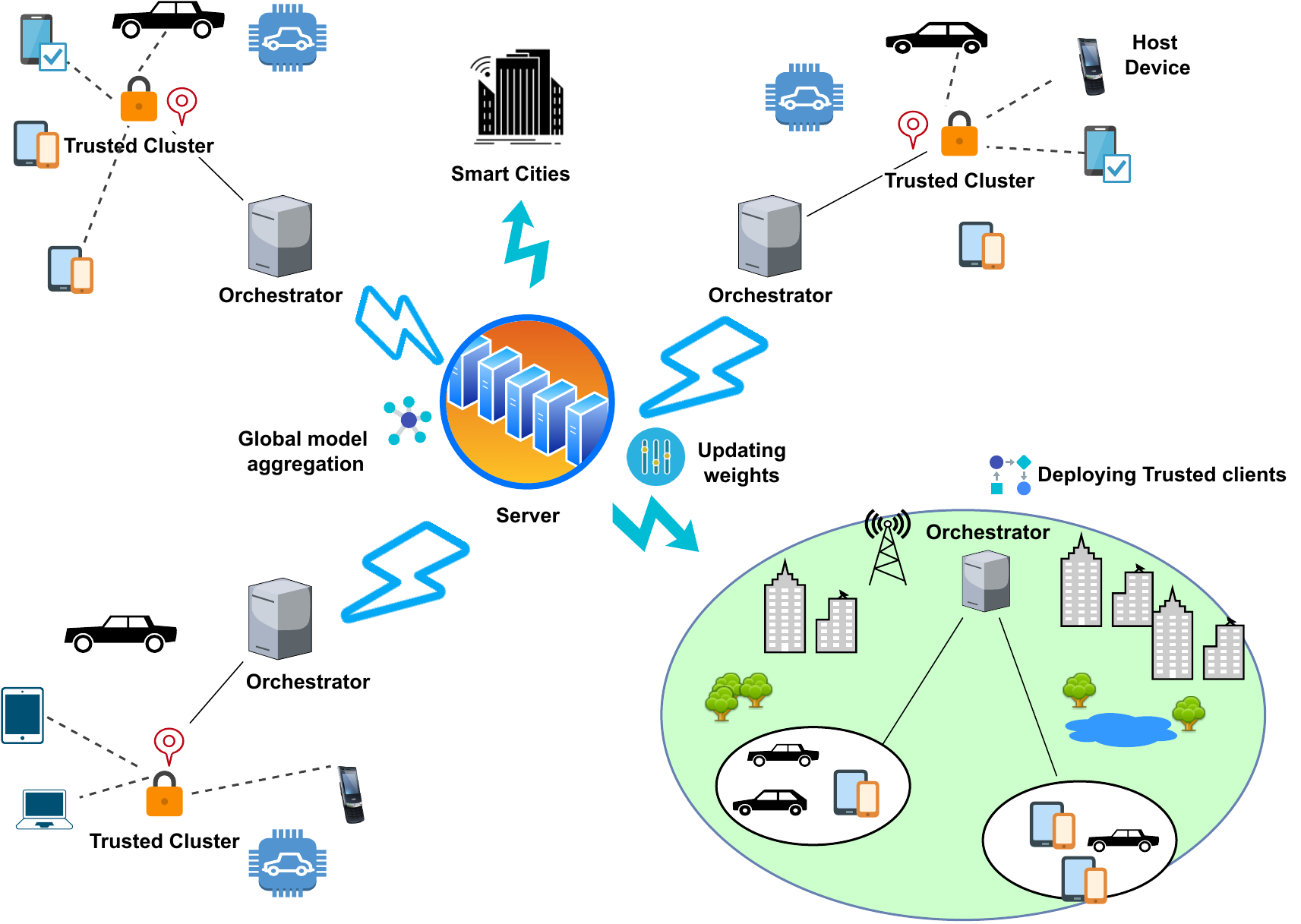}
	\caption{Overall Architecture.}
	\label{fig}
\end{figure*}

\subsection{Architecture Overview}

In our presented architecture \ref{fig}, the server manages container image creation and distribution, maintains the global model, ensures secure connectivity, and upholds cluster-wide trust. Orchestrators, on the other hand, handle cluster creation \cite{firstpaperr}, container deployment, device monitoring, client additions, and trustworthiness assessment.
Due to the adaptable architecture and streamlined containers, FL applications have the capability to execute an ML model at their convenience and in any location. This is made possible by leveraging any quantity of volunteered devices that engage as FL clients.
When an IoT device seeks to become part of a Kubeadm cluster, it might possess historical learning data, which the orchestrator can use to establish a trust rating. 
If there is no existing data, the Initial Trust Generator component is used to guarantee fairness and allocate an initial trust score to the newly joined clients.
Each device that joins the kubeadm cluster will have a log table for each client (its ID, device type and number, movements, successful number of assigned jobs, and trust level). Since we are deploying models and services on these devices, a client that was expelled out is unable to re-enter using a different ID.
After that, clients are assigned a level of trust, and it changes in every participating round.
The orchestrator and clients jointly strive to maintain a high trust level, which relies on a two-step verification process, their environmental conduct, and their behavior. This two-step verification serves the purpose of revealing any occurrences where clients introduce random weights or manipulate their labels, as malicious actions could potentially skew the global model in favor of specific class labels, thereby compromising the model's resilience and trustworthiness.
The actions of a client's environment and behavior can be used to derive the following numerical values: (1) the number of successful model deployments; (2) behavioral tendance to change the nature group that each client belongs to; and (3) a check for discrepancies in the context sharing tables of devices that the orchestrator has received from.

The orchestrator will deploy containers on compatible client devices. To put it another way, the container needs sufficient resources to complete a task, and host devices cannot allocate fewer resources since the container manager would raise a flag and not run.
Any significant variation from that period can raise questions about the device's willingness to share resources with the container. Additionally, given how dynamic and crowded today's smart cities are, some devices may contain more data from earlier rounds, which will make the model take longer to complete on the end device.
Our approach accounts for this by analyzing the average movements of the device. High movements result in high data generation rates.

Normally, data generated within a container can only be accessed within that container and solely while it's running. This data, including weights, is shielded from alterations. Host devices grant the container access to their data, and the container manages data division and preparation. Any host-level data changes can be identified during learning rounds but, once at the container level, data becomes inaccessible and unmodifiable by host devices.

Then, using genetic optimization, our architecture will deploy a set of clients while optimizing the number of clients deployed, their actions, their placement within an area, their level of trust, and the type of data they contain.


In our architecture, 
The Kubeadm Containerization Required Modules need to be operational across all nodes within our design, encompassing the Server, Orchestrator, and Fog nodes. The server hosts multiple components including the Aggregator, Client Deployment, Kubeadm Environment Initializer, Oracle Engine, and Communication Manager.
The orchestrator is employed to manage the components including Client Manager, Client Deployment, Initial Trust Generator, and Trust Detector. On the client nodes, the Fog Client, Learning Respond, and Context Sharing components are executed.

\subsection{Communication Manager}
This component manages communication tasks like establishing connections, broadcasting cluster invitation requests, and redistributing global weights among the nodes in the proposed architecture. Additionally, it enables further isolation through custom-defined bridges. When no network is specifically chosen, all containers are automatically linked to the bridge network. This feature allows unrelated containers to communicate, which may pose security risks. Communication is limited to containers connected to user-defined networks, creating a more controlled and scoped network environment.
\subsection{Aggregator}
The aggregator's role involves enhancing the global model using an FL aggregation function while monitoring the volume of received weight updates in each round. Furthermore, its duty includes verifying the appropriate threshold to omit a round, triggered by a situation where the count of updated weights goes below a threshold set by the server.
\subsection{Kubeadm Containerization Required Modules}
The Kubeadm tool enables the streamlined and secure establishment of Kubernetes clusters, a process made efficient by its integration with Docker, which is required to be operational on all devices consistently. An added advantage is Kubeadm's adaptability across various platforms. For communication management, cluster well-being upkeep, and device and service analysis, the deployment of Kubectl is necessary. Once client devices and the orchestrator/master node are set up, images are retrieved, initiated, and activated.
\subsection{Kubeadm Environment Initializer}
For the available volunteer devices within each region, the server initiates several orchestrators. The selection of initial orchestrators is determined using the recently gathered client device movement data. These orchestrators are subsequently tasked with fulfilling this role. 
In areas where multiple devices are enthusiastic to contribute, the server strives to prioritize the allocation of orchestrators. 
With this setup process complete, the environment is ready to facilitate on-demand services without any additional configuration time needed.
\subsection{Client Manager}
The orchestrator must first observe the motions of their clients.
When an orchestrator identifies significant mobility and a scarcity of reliable clients in a particular area, it will make a request to the server to deploy containers on a portion of these clients. The suggested framework avoids excessive duplication of client selection in subsequent rounds by monitoring the overall count of rounds in which a client has engaged. Additionally, this element supervises the remaining time a node spends within the cluster and assesses its average presence in a specific area, contributing to the achievement of the training objective. 
\subsection{Fog Client}
This module acts as the central channel for reporting details about clients’ profiles to the orchestrators, thereby fulfilling an important function in maintaining continuous updates for the orchestrators concerning any alterations or developments concerning these profiles. 
Its responsibilities include not just the initial submission of clients' profiles, but also the continuous task of swiftly communicating any subsequent updates that emerge. 
\subsection{Initial Trust Generator}
This module is responsible to assign initial trust levels for clients that do not have previous experience in the Kubeadm clusters. Having initial trust values will help the model converge faster, instead of starting from zero and building the trust from scratch \cite{osamaboost}. 
Bootstrapping is a method of recommendation employed across various sectors, such as cloud computing, to validate the reliability of recently deployed cloud services in cases where there exists no historical data regarding their past performances \cite{bosstr}. In our method, we will use bootstrapping to assign an initial trust value to the new devices that wish to join the clusters. The suggested technique relies on the collaboration of many active orchestrators to assist in overcoming the challenges by establishing initial trust for incoming devices. 
Every orchestrator maintains a collection of client device data from previous training iterations. This collection should encompass the attributes of each participating client device in the round (such as location, device category, and resource utilization), along with the level of trust earned by the device during the allocated round. When a new device enters the cluster with no prior trust, the server will anticipate the initial trust value by asking orchestrators to send over log datasets about their participating and reachable clients. 
Using the gathered datasets, a decision tree regression model is trained to forecast the anticipated trust value. By utilizing Standard Deviation Reduction (SDR) in place of Information Gain, the ID3 method becomes suitable for constructing a regression decision tree \cite{reftostd}. The standard deviation reduction approach is employed to evaluate the uniformity of a feature.

\subsection{Learning Respond}
This component is tasked with creating the learning service's dataset, which involves activities like retrieving data, performing data preprocessing, and dividing the data into training and testing subsets.
\subsection{Context Sharing}
Client devices have also a role in detecting malicious clients. This module is operated by the participating clients. It is responsible to share log files between neighboring clients in its area to detect any contradiction and false information. Moreover, the trust level of each neighboring device is shared between these clients so better strategies and influence they will have on other devices.
\subsection{Trust Detector}
This component assesses and consistently refreshes the trust level assigned to individual clients. Detailed presentation and implementation are found in Section IV.
\subsection{Module for Decision-Making in the Selection and Deployment of Trusted Volunteer Clients}
The orchestrators activate this module to effectively deploy clients within given regions. The effectiveness of client deployment is achieved by factoring in various criteria, constituting a multi-objective challenge. This particular issue is introduced and addressed in Sections IV and V.

\section{Problem formulation}
In this section, we present the multi-objective optimization problem concerning the deployment of models and clients, along with the Trust Detector module. 
\subsection{Problem Definition}
With a collection of accessible volunteer devices denoted as $D$ = \{$D_1$, $D_2$, ..., $D_n$\}, accompanied by their associated learning utilities $UT$ = \{$UT_1$, $UT_2$, ..., $UT_n$\}, our goal is to identify the suitable group of clients to engage in each learning iteration while considering various factors such as resource usage, priority, geographical location, availability, mobility, data characteristics, and trustworthiness. The task of selecting the optimal clients and effectively deploying them for learning is a challenging problem, known to be NP-hard. To simplify this problem, we model it as a knapsack problem and demonstrate its NP-hardness through proof.

\subsection{Components Formulation}

The aim of our optimization model is to reduce the count of deployed clients, increase the count of trusted clients, limit model control to certain class labels, enhance data diversity, and optimize the total number of trusted deployment requests.
\subsubsection{Input Data}

In the given problem scenario, we possess a collection of host devices denoted as $DC$, dispersed across different regions. Our goal involves deploying containers onto a subset of these hosts for participation in a learning iteration. The matrix representing the available hosts is denoted as $\boldsymbol{DC} \in \mathbb{R}^{n \times 6}$, where each host is characterized by six input attributes. These attributes encompass CPU resources (measured in Google compute units - GCUs), memory capacity (in bytes), disk space (in megabytes), battery level (as a percentage), availability (expressed as a numerical value), and geographical location within an area. \\

    \begin{center}
         $\bold{DC}_{i} = [DC_{i,CPU}, DC_{i,memory} ,  DC_{i,diskspace}, DC_{i,battery} ,  DC_{i,availability},$  $DC_{i,area}]$ $\forall i \in \{1, \dots, n\}$ where:
   
    $DC_{i,CPU}$ : CPU availability on $DC_{i}$.
    
    $DC_{i,memory}$ : Memory availability on $DC_{i}$.
    
     $DC_{i,diskspace}$ : Disks pace availability on $DC_{i}$.
    
     $DC_{i,battery}$ : Battery level on $DC_{i}$.
    
   $DC_{i,availability}$ : Availability time of $DC_{i}$ 

    $DC_{i,area}$ : Current area location of $DC_{i}$.
    
\end{center}
In addition to that, records of trust will be derived and integrated as input for the optimization.

Every host device in the set $\bold{DC}$, denoted as $\bold{DC}_{i}$, is associated with a utility input represented by a matrix $\bold{UT} \in \mathbb{R} ^ {n * 4}$. This matrix $\bold{UT}$ consists of four input features that indicate the CPU utilization, memory usage, battery consumption, and disk space utilization of the service running on the hosts.

  \begin{center}

    $UT_{i} = [UT_{i,CPU}, UT_{i,Memory}, UT_{i,Battery}, UT_{i,Diskspace}] $ where:

    $UT_{i,CPU}$ : CPU consumption of client $DC_{i}$.
    
    $UT_{i,Memory}$ : Memory consumption of client $DC_{i}$.
    
    $UT_{i,Battery}$ : Battery consumption of client $DC_{i}$.
    
    $UT_{i,Diskspace}$ : Disk space consumption of client $DC_{i}$.\\
      \end{center}
\subsubsection{Output Data}

 The set of clients selected to participate in a learning round is denoted as $\bold{O}$, which is the outcome of a multi-objective optimization model. The deployment of these clients is represented as a one-dimensional array $\bold{O}_{j}$, where $j$ represents the device number, $\forall j \in 1, .. , n$. Each element $\bold{O}_{j}$ of the array takes a value of either 0 or 1, indicating whether the corresponding host device $DC_{i}$ is chosen to be included in the learning round. A value of 1 for $\bold{O}_{j}$ indicates that $DC_{i}$ is selected, while a value of 0 indicates it is not chosen.
 
\subsubsection{Objective Functions}

Each objective function [$F_1$, $F_2$, $F_3$, $F_4$, $F_5$] is multiplied by its corresponding weight [$W_{f_1}$, $W_{f_2}$, $W_{f_3}$, $W_{f_4}$, $W_{f_5}$].

\subsubsection*{Reduce the count of clients that have been deployed}
    
The goal is to reduce the count of clients that are deployed during the learning phase.
    \begin{equation}
        F_1 = min(\sum_{i=1}^{n} O_{i})  \times W_{f_1}
    \end{equation}

    By minimizing the number of active hosts, we can achieve energy and resource savings, resulting in reduced battery, CPU, and memory consumption across all available clients. Consequently, this enhances the accessibility of fog devices, enabling additional services or applications to make use of the extra hosts within a specific region. Moreover, in the context of FL, both server and client devices encounter difficulties related to the substantial exchange of parameters and weight updates. By considering this objective, we can alleviate network congestion. As discussed in a previous study \cite{mario5}, there is no need to select the maximum number of clients for each round. Opting for a smaller number of clients in the initial stages and gradually increasing the number in subsequent rounds can enhance the training loss and ultimately lead to higher accuracy especially when we have trust values taken into consideration.\\

\subsubsection*{Maximize The Number Of Trusted Clients}
Clients with high Trust should be considered to participate in the learning. For that, this objective function tries to maximise the deployment of clients having high values of trust.
    \begin{equation}
        F_2 = max(\sum_{i=1}^{n} O_{i} \times Trust_{DC_i} ) \times W_{f_2}
    \end{equation}

Selecting clients with high levels of trust brings several advantages to the convergence of the model and contributes to a more efficient learning process. When clients with a proven track record of trustworthiness are chosen, the likelihood of encountering malicious attacks or unreliable behavior within their updates is significantly reduced. This is crucial because malicious updates can introduce noise or biased information that hinders the convergence of the model. The reduced presence of malicious or untrustworthy updates also safeguards the integrity and reliability of the overall learning process.\\

\subsubsection*{Minimize model controlling}

This objective function helps in limiting model tendency to specific class labels where compromised clients act normally in the learning stage but do some malicious actions on their side like hiding some data and forcing the model to learn on specific class labels more than others.
 
    \begin{equation}
        F_3 = max(\sum_{i=1}^{n} O_{i} \times RR ) \times W_{f_3} 
    \end{equation}
where 
$RR$ is the difference between the clustering rates of the selected clients.
    
A substantial $RR$ score suggests that the chosen clients originate from distinct clusters of data characteristics. Data produced by these clients exhibit dissimilarity, particularly in terms of the number of output classes and data volume. $RR$ is calculated by clustering clients based on their local accuracy. Choosing clients from different clusters help in having multiple data types and data nature while deploying clients \cite{proveclustering}. When devices have similar accuracy levels, it means they have comparable data distributions and common patterns. The integration of data from different clusters gives new views, allowing the model to capture a broader range of patterns, nuances, and edge cases \cite{proveclsutering2}. By that, the model is prevented from tending towards class labels more than others, which will make the model more robust.\\

\subsubsection*{Maximize The Diversity And Volume Of The Overall Data}

The purpose of this function is to enhance data volume and diversity.
  \begin{equation}
        F_4 = max(\sum_{i=1}^{n} O_{i} \times R ) \times W_{f_4}
    \end{equation}
where 
$R$ is the difference between the rates of the selected clients.
    
   A notable $R$ score indicates that the chosen clients come from distinct areas and exhibit considerable mobility. Data produced by such clients differ, particularly in terms of data nature.  
    Increased mobility generates a significant volume of data, enhancing the effectiveness of model training. The database of a user's device expands as they engage in frequent movement and utilize their device more frequently. This expansion is attributed to the storage of location-based data and usage data within the device's database, both of which gradually accumulate over time.
The presence of location-based data, such as GPS coordinates and timestamps, along with usage data, including app usage patterns and social media interactions, contributes to the growth of the device's database. These types of data are typically stored within the device and increase as users engage in activities that involve location tracking, navigation, location-based services, and social media usage. Consequently, individuals who exhibit a higher degree of mobility are more likely to utilize their phones for such purposes, leading to an even greater accumulation of data. thus, a more heterogeneous and wider range of data the model now can learn from. Moreover, the model can ease the issues created by changes in distribution by combining data from many area locations. The model trained on a broad dataset is better suited to dealing with variations in data patterns, making it more flexible to shifting conditions \cite{multiarea1}. Furthermore, it promotes transfer learning. The model may extract significant features and build transferable representations that generalize well across different locations by training on diverse datasets from different area locations \cite{multiarea2}. This allows for efficient information transfer and lowers the need to retrain models from the ground up for each new area location. 
A diversity in the output class in each round also helps the model to perform better on new testing data.\\

\subsubsection*{Maximize The Trusted Deployment}
     The objective is to maximize serving the requests sent from the orchestrators to deploy containers in some areas having few trusted clients.

    \begin{equation}
    F_5 = max(\sum_{i=1}^{n} O_{i} \times RT ) \times W_{f_5}    
    \end{equation}
where 
$RT$ is the average rate of clients selected from the requested areas by the orchestrators.
    
   The mini-servers or orchestrators play a crucial role in overseeing the availability of trusted clients within specific areas. Their primary responsibility is to keep track of the number of trusted clients present in each area. In situations where there is a shortage of trusted clients in a particular area, the mini-server takes proactive action by sending a request to the central server for container deployment.

Upon receiving the request, the server evaluates the situation and initiates the deployment of containers. The containers are strategically deployed on clients located in the requested areas. This deployment strategy ensures that the learning process can effectively utilize the available trusted clients in the desired regions.

By dynamically deploying containers in areas with a limited number of trusted clients, the system aims to address the imbalance in client availability. This approach ensures that the learning process can be conducted efficiently across different areas, maximizing the utilization of trusted clients and optimizing the overall performance of the FL system.

\subsubsection{Trust factor}
Our Trust factor of each client is formulated by aggregating four components: 
\begin{enumerate}
    \item  Number of successfully serving deployed jobs - Algorithm \ref{alg:cap1}:
        Organizations can assess the ability of the device to continuously execute deployed tasks without errors or disruptions by recording the number of tasks done successfully. This metric allows for the detection of probable anomalies or deviations from typical behavior, which may indicate potential security breaches or failures \cite{form1}. Devices with a high percentage of failed tasks, on the other hand, may raise worries about their dependability, performance, or potential vulnerabilities.
        
        This measurement aids the orchestrator in keeping track of how many client tasks have been successfully completed. Due to the design of our architecture, we can deploy containers and volunteer devices that have expressed an interest in joining the cluster. However, some devices might not fully collaborate once containers have been deployed on such devices while taking into account the constraints and limitations. Certain devices may hide specific local log files when it comes to context sharing, while others may drop rounds not due to network issues or packet loss but as a subtle attempt to delay by consuming more shared resources with the containers, leading to their exclusion from the round. By factoring in these suspicious behaviors, we can develop a more comprehensive assessment of the trustworthiness of client $i$.

    \begin{algorithm}
\caption{Number of successfully serving deployed tasks}\label{alg:cap1}
\begin{algorithmic}
\State $Tr_1 \gets 0$
\State $success$ : count of the successful deployed jobs of $DC_i$
\State $deployed$ : count of the deployed jobs of $DC_i$
\State

\If {$deployed \geq$  1}

\State{$Tr_1$ = $\frac {success}{deployed}$}
\EndIf

\State \textbf{Return $Tr_1$}
\end{algorithmic}
\end{algorithm}

    \item Two-step verification - Algorithm \ref{alg:cap2}:
         
        In this interactive process, both the orchestrator and individual clients participate in a game to collect information concerning their class labels and any potential label manipulations. Using their respective local test data and relaying local accuracy measurements to the orchestrator, the orchestrator initially focuses on clients with lower trust levels in the initial rounds. Rather than sending these clients the most recent global weights, the orchestrator transmits older global weights for comparison with their local accuracy. Suspicious clients frequently engage in label flipping and data label modifications to reduce the model's reliability and introduce deception. 
        Issues concerning label manipulation by clients, Inconsistencies in local accuracies despite using identical prior global weights, and the transmission of incorrect local accuracy data to sabotage the global model may arise (Forman, 2016). These actions exemplify potential occurrences during learning, such as concealing or altering labels to mislead the model; clients engaged in such behavior should face repercussions through a reduction in their Trust score.

        In our dynamic environment, it's possible for clients to have varying amounts of testing data from one round to another. Given that prior weights were computed from a limited number of recent rounds, minor fluctuations in local accuracy are tolerable, considering the fluctuating data volume. To detect unusual behavior in terms of local accuracies among IoT devices, the model employs a modified Z-score statistical technique, as described in \cite{lit7}. Unlike the regular Z-score, the modified version uses the median instead of the mean, rendering it less sensitive to outliers. This method calculates the score by measuring the difference between local accuracies in a single round and the median accuracy from previous rounds and then dividing it by the median absolute deviation (MAD) of the metric, denoted as z. According to \cite{inalgo}, values with modified z-scores greater than 3.5 or less than -3.5 are considered potential outliers. However, the server has the flexibility to adjust this parameter based on its historical experience in specific areas.


    \begin{algorithm}
\caption{Two-step verification}\label{alg:cap2}
\begin{algorithmic}
\State $Tr_2 \gets 0$
\State $j$ : is the Client device that is monitored
\State $Ac$ : is the Accuracy level to be analyzed by the orchastrators
\State $M_j^z$ : a table recording the $x_j^z$ 's accuracy recorded in each round.
\State $\bar{x_j^z}$ : is the median of $M_j^z$
\State $MAD_j^z$ : the median absolute deviation of $M_j^z$. i.e: $MAD_j^z$ = median \{ $|x_j^z$ - $\bar{x_j^z}|$ \}
\State $\alpha_j^z$ : the modified Z-score of $x_j^z$
\State $Counter\_abnormal\_metric_j^z$ : number of abnormal Accuracy occurrences by j
\State $Counter\_abnormal\_metric_j^z\_average$ : number of abnormal Accuracy occurrences by j
\State $\epsilon$ : Threshold to determine an outlier.
\State
\State $Counter\_abnormal\_metric_j^z$ $\gets$ 0
\State Calculate the median $\bar{x_j^z}$ of $M_j^z$
\State Calculate the $MAD_j^z$ of $M_j^z$
\State Calculate $\alpha_j^z$ = $\frac{0.6745 (x_j^z - \bar{x_j^z})}{MAD_j^z}$

\For{\textbf{each} data point $x_j^z \in$ $M_j^z$  }  
\If {$\alpha_j^z \geq$ $\epsilon$ }
\State $Counter\_abnormal\_metric_j^z\_average$ += ${x_j^z}$ 
\State $Counter\_abnormal\_metric_j^z$ += 1
\EndIf
\EndFor

\State $Tr_2$ = $Counter\_abnormal\_metric_j^z$ 
\State
\State \textbf{Return} $Tr_2$
\end{algorithmic}
\end{algorithm}

    \item Large deviation from the expected group - Algorithm \ref{alg:cap3}:
        Our architecture determines whether the selected client has sufficient resources before deploying the model to a device. The training data and data volume of each device are not available to us. Environmental activities, though, can provide us with a clue. Environments with highly dynamic devices and a high level of movement can produce a lot of data. When compared to clients from other places, clients from this area typically take longer to complete their jobs. To do this, we group participating clients into clusters according to their Type, Average Movements, Average Round Finish Time, and CPU and RAM Capabilities \cite{form3}. By clustering clients according to similar data volumes generated from their motions and average task completion times while taking into account their resource capacities, we can efficiently group these devices. If a client is performing its tasks as usual while learning, it will most likely remain in the same cluster of the few previous rounds. However, if the orchestrator observes a client $i$ frequently changing its group, this raises questions about its dependability of completing before or after the anticipated time to finish while it is grouped based on its data volume (movement), its resource capabilities, and its past completion times especially while having containers deployed on these clients. Clients who change clusters frequently or display unusual behavior may raise suspicions and necessitate further investigation to protect the integrity of the FL process.

        The algorithm consists of clustering the clients based on the above-mentioned criteria using agglomerative clustering \cite{form6}. It captures both local and global commonalities among clients. By iteratively merging clients based on similarity measurements, it promotes the creation of clusters that enclose similar patterns inside themselves. By using an incremental technique, agglomerative clustering may handle large-scale FL scenarios. Instead of processing the full client population at once, it can gradually combine clients, making it suited for resource-constrained environments and tolerating constantly changing client populations \cite{form5}. The large deviation is then estimated by comparing the number of common neighbors in this round to the previous round. This will give an idea of whether or not the client is sticking with the same group of clients who have similar resource capabilities, mobility (volume of data), and average finishing time.

        

    \begin{algorithm}
\caption{Large deviation from expected group}\label{alg:cap3}
\begin{algorithmic}
\State $Tr_3 \gets 0$
\State $NormalMatrix$ : the normal behaviour of a client.
\State $Groups$ = Cluster ($NormalMatrix$)
\State $DC_i$.add(new group members)
\For{\textbf{each} new neighbor $N_i$ of $DC_i$}  
\If {$check\_common\_neighbor(N_i)$ == True}
\State $Tr_3$ +=1
\EndIf

\EndFor
\State \textbf{Return} $Tr_3$

\State
\Function{check\_common\_neighbor}{$N_i$}
\For{\textbf{each} normal neighbor $No_i$ of $DC_i$}:
\If {$No_i$ == $N_i$}
\State \textbf{Return} True
\EndIf

\EndFor
\State
\textbf{Return} False
\EndFunction


\end{algorithmic}
\end{algorithm}
    
    \item Context sharing contradiction - Algorithm \ref{alg:cap4}:
     In our architecture, the deployed clients have the security responsibility of surveying and reporting. Each client will be gathering shared data from its local neighbors. Deployed clients will submit their local log file and the most recent weights for each round. Any discrepancy between the data the orchestrator receives from clients will be checked for by the orchestrator. A client's data with a high number of contradictions will raise greater questions about its reliability \cite{form4}.
        
\end{enumerate}

    \begin{algorithm}
\caption{Context sharing contradiction}\label{alg:cap4}
\begin{algorithmic}
\State $Tr_4 \gets 0$
\State $K_i$ : the list of values reported by the client
\State $Ko_i$ : is the list of values recorded by the orchestrator
\State $count\_contradiction$ : count the number of contradiction by client $DC_i$

\For{\textbf{each} metric value $s$ in $K_i$}
\State $So$ = metric $k$ in $Ko_i$
\If {$s$ != $So$}
\State $count\_contradiction$ +=1
\EndIf
$k$ = $k$ +1
\EndFor
\State $Tr_4$ = $count\_contradiction$
\State
\State \textbf{Return} $Tr_4$ 
\end{algorithmic}
\end{algorithm}

Our trust factor is added to the log file of the orchestrators after aggregating the scaled four measurements  (\ref{eq}) in every learning round.

\begin{equation}
\label{eq}
    Trust = \dfrac {(Tr_1 \times \alpha_1) - (Tr_2 \times \alpha_2) + (Tr_3 \times \alpha_3) - (Tr_4 \times \alpha_4)} {4} 
\end{equation}
where the $Trust$ Value is between 0 and 1. If the client does not show any violation of the above measurements will have a value 1 as a full trust, else it will be a number between 0 and 1. The $\alpha$ sign is an additional weight to each measurement decided by the orchestrators in case suspicious attacks are more common to happen in an area. \\


\subsubsection{Constraints}
 \hfill \\
 \textbf{Resource constraints:} 

 A client, denoted as $DC_{i}$, is considered for selection if its resource capabilities are not fully utilized upon deployment. The hosting capacity of the client is determined by factors such as CPU, memory, disk space, and battery level. Additionally, the learning utilization, which includes CPU, memory, disk space, and battery consumption, is taken into account.
 This constraint ensures that clients are selected based on their remaining capacity to accommodate the learning workload. By considering both the resource capabilities and utilization, the system can optimize the allocation of clients, maximizing the utilization of available resources while preventing the overloading of individual clients.

    \begin{equation}
    UT_{i,CPU} \times O_{i} \le DC_{i,CPU} 
    \end{equation}
    \begin{equation}
    UT_{i,memory} \times O_{i} \le DC_{i,memory} 
    \end{equation}
    \begin{equation}
    UT_{i,diskspace} \times O_{i} \le DC_{i,diskspace}
    \end{equation}
    \begin{equation}
    UT_{i,Battery} \times O_{i} \le DC_{i,Battery}
    \end{equation}
    
    $\forall i \in \{1, \dots, n\}$ i.e. for all available clients $DC_{i}$ and their utilization $UT_{i}$.\\
    
    \textbf{Minimum availability time:}
    The purpose of this is to prevent devices from exiting a learning round because of their mobility and the duration they remain within an area or cluster. To achieve this, the server determines a parameter denoted as ${T \in \mathbb{N}}$, which signifies the minimum required time for completing a single round. Therefore, a client can be deployed and selected if $DC_{availability_i}$, which is the staying time of $DC_{i}$ in its area, is greater than $ST$.
 \begin{equation}
    \forall i \in \{1, \dots, n\} \;\;\;\;\; DC_{availability_i} \ge ST
    \end{equation}
    
    
    \textbf{Optimized Trust deployment:}
    In our architecture, the measurement $Trust$ is taken into consideration while deploying trusted clients in each area. However, newly added clients will tend to have lower trust levels than experienced clients. For that, applying a balance in the way of deploying the clients by taking into consideration choosing some clients with low initial levels of trust. Without these constraints, the multi-objective optimization problem will always choose clients with the highest trust level neglecting the newly joined ones. 
    Therefore, in order to prevent redundancy during deployment, the server establishes a threshold to specify the proportion of clients with substantial trust. A client is deemed highly trustworthy if their $Trust{i}$ surpasses $MaxT$, where $MaxT$ is the designated threshold for identifying highly trusted clients. The same logic is applied in our algorithm for clients with high movements.
    
    \begin{equation}
        \sum_{i=1}^{n} O_{i} \le Mt \; 
        \forall i \in 1, .. , n, \; Trust_{i} \ge MaxT
    \end{equation}
    where $Mt$ is the portion of clients to be selected with high level of trust.\\
    
    \textbf{Weights summation:} To introduce greater adaptability according to the importance of the objective functions, the technique of adaptable weights \cite{hihi} proves beneficial. This involves multiplying each objective by a fractional value ranging from 0 to 1 (\ref{eq:1}), wherein their collective sum equals 1.
    \begin{equation}\label{eq:1}
        W_{f1} + W_{f2}+ W_{f3}+ W_{f4}+ W_{f5} = 1
    \end{equation}

In the problem of a single knapsack \cite{knapsack}, there are multiple objects available, each with its own weight and value. Alongside these objects, we have the capacity constraint of the knapsack. The main objective is to carefully choose the objects in order to maximize the overall profit while ensuring that the total weight does not exceed the capacity of the knapsack.

In this section, we encounter a multi-objective optimization problem that shares similarities with the well-known Knapsack Problem. Specifically, we can establish the NP-Hardness of the multi-objective optimization problem by reducing it to the multi-objective knapsack problem, as demonstrated below.

\textbf {Proof}: In the context of a client deployment problem instance, we simplify it into a knapsack scenario using the following approach: \\
\begin{itemize}
    \item Clients are treated as items within a knapsack, and the attributes associated with each client, including their Trustworthiness, mobility, availability, and resource capacity, are provided to the objective functions, aiding in quantifying the importance of a client (value of an item).
    \item The weights of an item are represented as the selection cost. This means a client must satisfy conditions such as having sufficient time to complete a round ($DC_{availability_i} \ge ST$), specific movement patterns, and resource utilization.
\end{itemize}
 
Conclusively, the aim is to optimize the utility (profit) obtained from the chosen clients while adhering to the constraints.

This reduction yields that our client deployment problem is NP-Hard.

\section{GENETIC ALGORITHM FOR TRUSTED CLIENT DEPLOYMENT}

Multi-objective optimization problems exhibit multiple solutions rather than a single optimal solution. These solutions are known as Pareto solutions, representing a trade-off between conflicting objectives. Obtaining the Pareto set of solutions efficiently and within a short timeframe is of utmost importance in such problems. To tackle this challenge, employing a Genetic Algorithm (GA) is a favorable approach \cite{GA}.

GAs are renowned for their evolutionary strategy, which emulates the process of natural selection. They select the fittest set of solutions to propagate and produce the next generations. 
In our genetic algorithm solution, a matrix $Z$ represents each chromosome. Each chromosome, designated as $Z_{j}$, symbolizes the decision made by the optimization model concerning the deployment of a particular client, $Z_{j}$, for a given service.
A value of 1 in $Z_{j}$ indicates that client $Z_{j}$ is chosen for deployment, while a value of 0 signifies that it is not selected. Thus, $Z_{j}$ takes a value within the range of $\in$ [0 , 1].

By utilizing a GA, we can effectively explore the solution space and identify Pareto-optimal solutions for the multi-objective optimization problem. The evolutionary nature of the GA enables the model to adapt and evolve over generations, iteratively improving the set of solutions and converging towards the Pareto front. 

\begin{algorithm}
\caption{Multi-objective genetic algorithm}\label{alg:cap6}
 \textbf{Data:} Set of available clients
 
 \textbf{Result:} Pareto set approximation $P_{known}$
\begin{enumerate}
    \item Check if the problem has a solution
    \item Initialize set of solutions $P_{0}$
    \item $P'_{0}$ = repair infeasible solutions of $P_{0}$
    \item Update set of non-dominated solutions $P_{known}$ from $P'_{0}$
    \item $Y \gets 0$
    \item $P_{X} \gets P'_{0}$
    \end{enumerate}
\begin{algorithmic}
    \While{(stopping criterion is not met),}
    \begin{enumerate}
    \setcounter{enumi}{6}
         \item $M_{X}$ = selection of solutions from $P_{X}\cup P_{known}$
         \item $M'_{X}$ = crossover and mutation of solutions of $M_{X}$ 
         \item $M''_{X}$ = repair infeasible solutions of $M'_{X}$ 
         \item increase Y
         \item Upgrade the solutions $P_{known}$ based on $M''_{X}$
         \item $P_{X}$ = fitness selection based on $P_{X}\cup M''_{X}$
    \end{enumerate}
    \EndWhile

    \State\textbf{Return:} Pareto set $P_{known}$

\end{algorithmic}
\end{algorithm}

Algorithm \ref{alg:cap6} outlines the implementation of the Genetic Algorithm (GA) in our solution. The initial stage entails assessing the possibility of deploying containers on accessible client devices. Following this, an initial set of solutions, labeled as $P_0$, is generated by picking clients for deployment randomly. The resulting solution is then assessed and rectified to correct any breaches of the stipulated constraints. The reparation process is described in Algorithm \ref{alg:cap7}, where containers are moved to other client devices under certain conditions.

\begin{figure}[]
	\centering
	\includegraphics[width=0.5\textwidth]{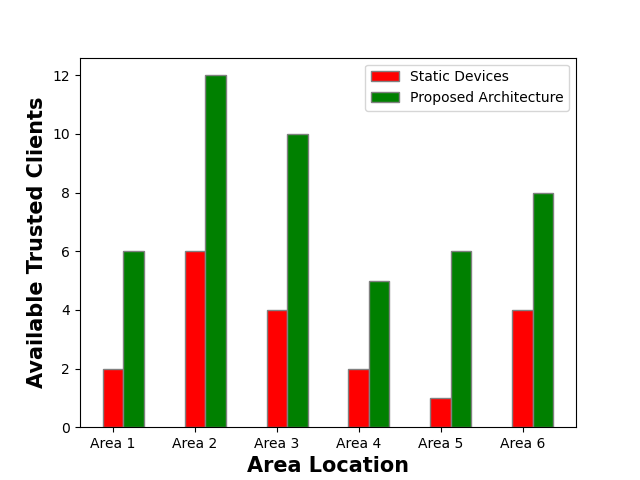}
	\caption{A snapshot on the number of available trusted devices in different area locations at a time t.}
	\label{g1}
\end{figure}

The corrective action is initiated when the ML process surpasses either (1) a client's resource capacity or (2) the predetermined duration for which the client is expected to remain in a specific location, or (3) thresholds defined as the "high movement" / "high trust" limits determined by the server. By resolving these violations, the algorithm ensures that the constraints are adhered to, maintaining the integrity of the solution.

Later, non-dominated or Pareto solutions are produced. Following this, the algorithm employs standard selection, crossover, and mutation operations. Unfeasible solutions are rectified, and updates are made to the Pareto set whenever enhancements can be made. This iterative procedure is reiterated over numerous iterations to enhance and optimize the solutions.

In our GA algorithm, binary tournament selection is used to choose individuals from the populations. One-point crossover is employed as the crossover method, with crossover applied to each parent based on a hyper-parameter-defined probability. For the mutation step, the bit string mutation technique is adopted, assigning each gene a mutation probability of 1/L, where L denotes the number of deployable clients. By integrating this mutation strategy, the algorithm avoids local optima traps, enhances diversity, and ensures fair opportunities for clients with lower mutation probabilities to undergo mutation.

Overall, this GA algorithm design aims to enhance the effectiveness of the solution, maintain diversity, and prevent premature convergence, ultimately improving the performance and quality of the obtained solutions.

The time complexity of this algorithm can be segmented into several components. Initially, we consider the overall count of the populations. Then, there's the initialization of the initial population, the establishment of constraints, the computation of fitness, the selection of parents, the application of crossover, and finally, the mutation phase. 
Let $CH$ and $NO$ be the number of chromosomes, and the number of nodes respectively.
Therefore, the overall complexity time will be:
O($CH \times NO$ + $C_{fitness}$ + $C_{fix-violation}$)).

\begin{algorithm}
\caption{Reparation Algorithm}\label{alg:cap7}
 \textbf{Data:} Infeasible Solution $L$, $n$ is number of elements in $L$
 
 \textbf{Result:} Accepted Solution
\begin{algorithmic}
\State accepted = False; $j$ = 0
\While{$j \le n$ and accepted == False} 
\If{$L_{j}$ is overloaded} 
\State deploy the container on a different capable client
\State in the same area location
\EndIf

\If{$L_{j}.availability \le threshold $} 
\State Place the container onto an alternate capable client
\State with a noteworthy history of frequent movements.
\EndIf

\If{$NN$ (highly movement clients)$ \ge threshold $}
\While{$NN \ge threshold $}   
\State swap (high move $DC$, not high move $DC$)
\EndWhile
\EndIf

\If{$HT$ (highly trusted clients)$ \ge threshold $}
\While{$HT \ge threshold $}   
\State swap (high Trust $DC$, not high Trust $DC$)
\EndWhile

\EndIf
\State $j$++

\EndWhile
\State\textbf{Return:} Accepted solution
\end{algorithmic}
\end{algorithm}

\section{Results and analysis}
To showcase the relevance and feasibility of our suggested approach, we employ a use case involving the prediction of subsequent locations in an application. This involves the utilization of an ML model to forecast the user's next visited location.

\begin{figure}[]
	\centering
	\includegraphics[width=0.5\textwidth]{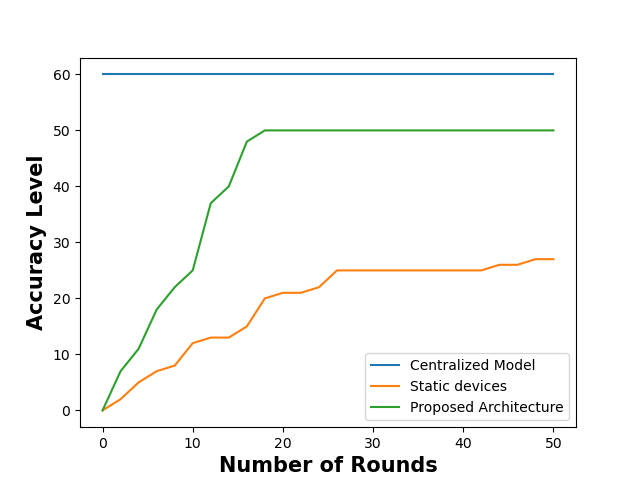}
	\caption{Accuracy progress of the centralized model, static devices architecture, and our proposed solution in relation to the progression of learning rounds.}
	\label{g2}
\end{figure}

In our training, the "MOBILE DATA CHALLENGE (MDC)" \cite{mario14} dataset is utilized. Given that we are dealing with datasets from the actual world. 
The MDC dataset comprises a significant volume of continuous mobile phone data employed for modeling individual behavior and social network dynamics.
We employed Python to execute the data preprocessing phase, which involved handling a dataset comprising continuous device movement records. By consolidating GPS coordinates, we grouped them into twenty distinct locations and six broader areas. The "area" feature represents a cluster of multiple specific places. Furthermore, we extracted and incorporated features such as date, day of the week, visit duration, and visit frequency for utilization in our model. 
The dataset for our 50 clients contains varying numbers of records, ranging from 200 to 1,500 entries. This data partitioning follows a non-identically distributed (non-IId) data distribution pattern, meaning that each client's data segments and class labels differ. We determine the allocation of user data to a specific client device based on the device's location. 
The simulations were conducted using the ModularFed framework \cite{arafehlocalfedd}.

As the base model, a 128-by-256 deep neural network with three hidden layers and the "ReLU" activation function is employed. 
The centralized base model attains a 60\% accuracy rate on the dataset, whereas our proposed FL model achieves 50\% accuracy. This performance disparity between the base model and the FL model is expected when working with real-world datasets and maintaining data privacy.

We examine situations where numerous volunteer devices are present in specific locations, necessitating the dynamic deployment of trusted clients in these areas. In scenarios where data and trusted clients are scarce, our architecture plays a pivotal role in facilitating the deployment and maintenance of a substantial number of trusted clients during the learning phase. We compare these simulations with the default random selection method \cite{mario12}, the centralized model and the trust establishment approach described in \cite{lit7}, named DDQN-Trust. 
As pioneers in introducing a trust mechanism for the On-Demand architecture, our comparison extends beyond the conventional FL malicious client detection methods. Unlike previous works such as \cite{lit7}, which incorporated trust factors solely during client selection, we integrate trust considerations throughout both the deployment and client selection phases in our On-Demand architecture. While \cite{lit7} took into consideration a trust factor while selecting clients, it's essential to note that our approach to building the trust factor differs significantly, prompting us to benchmark against these established methodologies.

\begin{figure}[]
	\centering
	\includegraphics[width=0.5\textwidth]{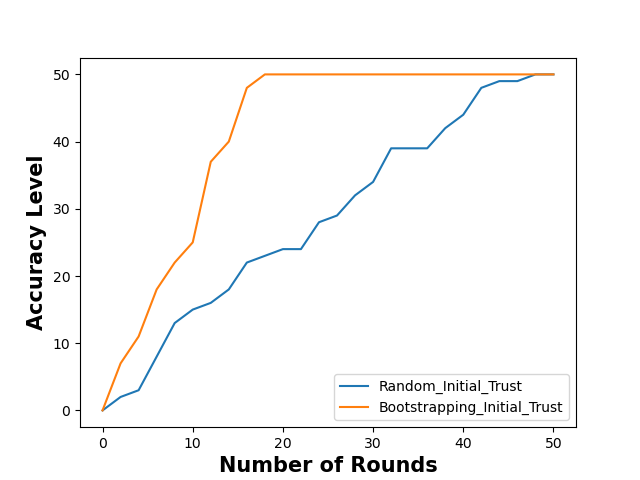}
	\caption{The Accuracy Levels of Trusted-On-Demand-Fl while using random trust initialization vs bootstrapping.}
	\label{g5}
\end{figure}

1) Figure \ref{g1} presents a compelling comparison between the quantity of accessible trusted clients and the reliance on static, trustworthy clients typical in traditional learning frameworks. The expanded pool of trustworthy clients in various locations presents a distinct advantage, as it amplifies the availability of reliable data sources. Conversely, environments dependent on static clients encounter challenges such as prolonged round-trip times due to distant clients and scarcity of trustworthy options, leading to frequent discards during the learning process.
Our innovative architecture effectively addresses these challenges by facilitating the deployment of trustworthy clients precisely where they're needed, even in remote or inaccessible areas. In the depicted scenario, the increased availability of trusted devices stands ready to contribute as FL clients, exemplifying the adaptability and scalability of our approach.
Moreover, Figure \ref{g1} showcases the continual growth of data volume available for learning, paralleled by the strategic deployment of volunteer clients. As the data reservoir expands, the role of orchestrators becomes increasingly pivotal. They oversee client movements and visited areas, ensuring optimal utilization of resources and maximizing the effectiveness of the learning process.

\begin{figure*}[]
	\centering
	\includegraphics[width=0.98\textwidth]{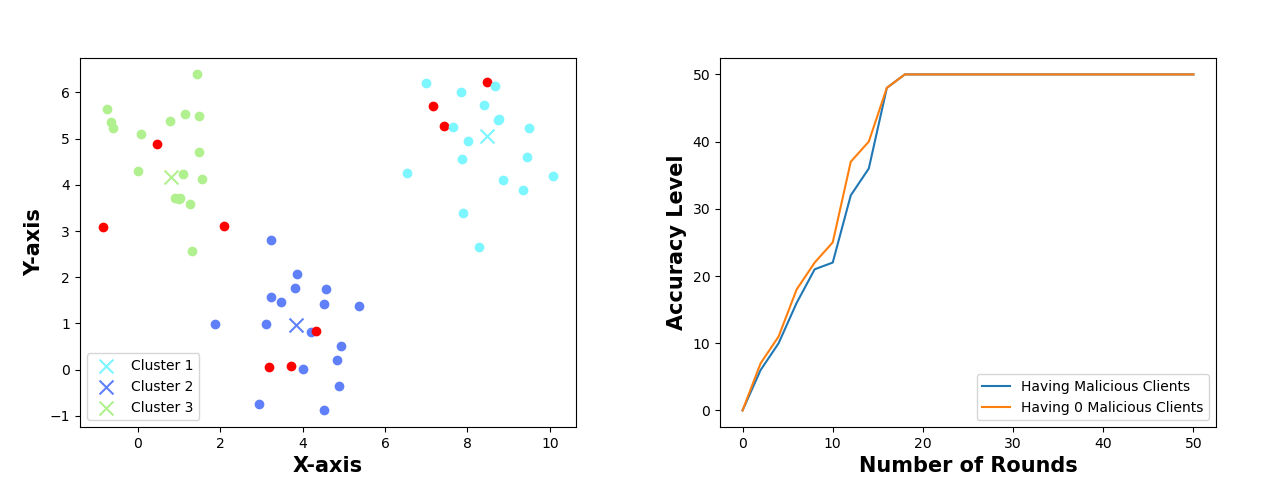}
	\caption{This graph shows the trend of certain devices deviating from their typical clusters while assessing the performance of our solution with and without these devices.}
	\label{g8}
\end{figure*}

2) In Figure \ref{g2}, we observe a progressive deployment strategy starting without any initially trusted clients. In each round, a fraction of clients is deployed and their trust levels are continuously monitored, resulting in a notable improvement in accuracy from zero learning to a substantial level with each passing round. This On-Demand container deployment approach, coupled with trust monitoring, underscores the significance of our proposed method.
Furthermore, when employing the conventional random selection method without considering trust and the above-mentioned objective functions, a significant number of rounds are discarded, causing delays and hindering the global model's quest for the desired accuracy, especially when dealing with untrusted actions from malicious clients. Conversely, by recording and leveraging the trust levels of each client and selecting those with higher trust, we achieve the desired accuracy with fewer discarded rounds.

3) The implementation of a trust bootstrapping mechanism is critical in the FL process. Several advantages are gained by continuously updating the trust factor and deploying devices based on the established objective functions.
Figure \ref{g5} shows a comparison of the accuracy levels obtained when using random trust initialization against the bootstrapping method. The graph clearly shows the significant gap between convergence time and trust value evolution. When the random assignment is used, the model has a long convergence phase, requiring numerous rounds of clients to progressively increase their trust levels. The bootstrapping technique, on the other hand, allows for the assignment of fair and appropriate beginning trust values. This increases the model's trustworthiness by allowing it to quickly access a larger pool of highly trusted clients. As a result, the model reaches convergence by round 20, whereas random assignment delays the process due to the initially low trust values assigned.
The trust bootstrapping technique is significant because of its capacity to accelerate the integration of trustworthy devices into the FL process. By starting with greater trust values, the model may better use the knowledge and contributions of trustworthy clients, resulting in increased performance and faster convergence.

4) Furthermore, throughout the execution of our FL application, we encountered instances where certain clients exhibited malicious behavior. Figure \ref{g8} illustrates a comparison between our solution with no malicious clients and our solution with added malicious clients. During the learning process, we intentionally introduced clients engaging in malicious activities, such as concealing their data from containers post-deployment and altering their behavior, specifically targeting the criteria outlined in Algorithm 3, which focuses on detecting large deviations from expected group behavior.
These malicious clients, exhibiting frequent cluster changes or unusual behavior, raise suspicions. Such anomalies occur when clients complete tasks before their expected time, based on their historical data volumes, resources, capabilities, and typical completion times. Any deviation from these patterns can indicate suspicious behavior.
In the left portion of Figure \ref{g8}, we observe three devices highlighted in red consistently changing clusters across multiple rounds. These devices either fail to process all available data, finish well before or after expected, or exhibit inconsistent behavior. Consequently, these devices receive lower trust scores. However, despite the presence of malicious clients, our solution demonstrates resilience by continuing to deploy and select high-quality clients, maintaining a high level of trust, learning accuracy, and performance compared to scenarios without malicious clients.

\begin{figure}[]
	\centering
	\includegraphics[width=0.5\textwidth]{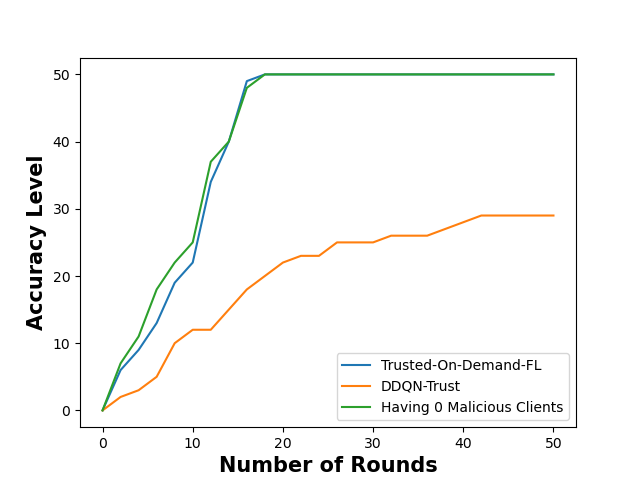}
	\caption{This graph shows a comparison between our Trusted-On-Demand-FL solution and the DDQN-Trust model while having clients acting maliciously.}
	\label{g9}
\end{figure}

5) Figure \ref{g9} illustrates the performance of DDQN-Trust \cite{lit7} in comparison to our approach, both with and without malicious acts. While we do not directly compare their use of Deep Reinforcement Learning for client selection, our focus lies in comparing their method of calculating Trust values, which inform their selection algorithm in each round. Since relying solely on objective functions for client selection is insufficient in deployment scenarios, we utilize their Trust mechanism for comparison.
Our architecture demonstrates the ability to detect malicious clients and assign fair Trust values to penalize their actions. The graph showcases the impact of identifying instances where clients attempt to manipulate their training data labels, intentionally undermining the model's performance. Our Trust factor addresses this issue by evaluating testing accuracies based on previously calculated weights. As depicted, our approach consistently maintains a highly trusted environment with high accuracy. In contrast, DDQN-Trust struggles to achieve satisfactory performance, especially in scenarios where clients constantly change and flip labels, reaching a maximum of 30\% accuracy. If we solely rely on monitoring resource utilization to detect suspicious behavior, these clients will persist in manipulating labels while consuming resources at normal levels, hindering model progression and convergence.
Furthermore, our objective functions play a pivotal role in early accuracy enhancement by selecting clients from diverse geographical locations and different clusters. This strategy ensures greater data heterogeneity, enriching the model's learning experience. By combining these approaches, we evaluate client behavior within the FL system and assign unique trust ratings accordingly. This evaluation extends beyond mere resource monitoring \cite{lit7} to encompass container-level considerations, where devices may utilize resources for other applications. This approach enables us to prioritize resources for the most reliable clients while mitigating the influence of malicious entities through appropriate penalization.


The ability to recognize malicious clients and assign distinct trust values not only helps safeguard the integrity of the FL system but also ensures a fair and equitable allocation of resources. By accurately identifying and addressing the presence of malicious behavior, our architecture promotes a more secure and reliable learning environment for all participants involved.
The above graphs serve as clear evidence of how our approach accounts for attempts to undermine the model's integrity by considering label flipping. Through these mechanisms, our architecture ensures a robust trust evaluation process, leading to improved model performance and reliability. This approach promotes enhanced diversity in the data utilized for training the model. The inclusion of diverse data sources contributes to a more comprehensive and representative learning experience. Additionally, the technique facilitates an increased volume of data, thereby enriching the model's knowledge base.
 




\section{Conclusion}
The suggested architecture investigates FL client availability, reliability, and dependability. 
Moreover, it facilitates the deployment of clients while simultaneously evaluating their reliability through the calculation of a trust score for each. The trust score undergoes continuous evaluation and modification, taking into account factors such as the two-step verification process, the number of successfully completed deployment tasks by each client, significant deviations from expected behaviors, and discrepancies in the log files. Leveraging Kubeadm and containerization advancements, we efficiently manage dynamic deployment, ensuring that clients are accessible on resource-constrained devices at any time and location. By leveraging a GA, we introduce a multi-objective optimization approach to resolve the deployment of both models and clients. The experimental results show that the suggested approach's efficiency and upgrades are effective since fewer rounds are required to achieve the required accuracy by eliminating untrustworthy and unreliable clients. Furthermore, more dependable and trustworthy clients are available and assessed for learning in a variety of areas. In the future, we intend to develop a deep reinforcement solution for the deployment process, as well as other prediction approaches for Trust factors, area placements, and deployment.

\printcredits

\bibliographystyle{cas-model2-names}

\bibliography{cas-refs}


\bio{mario}
Mario Chahoud received his M.Sc. degree and B.S. degree in Computer Science from the Lebanese American University (LAU). He is currently a Research Fellow at the LAU Cyber Security Systems and Applied Artificial Intelligence Research Center and at Mohamad Bin Zayed University of Artificial Intelligence (MBZUAI), Abu Dhabi, United Arab Emirates. He was a Research and Teaching Assistant at the Lebanese American University. He is currently a Ph.D. student at Concordia university, Montreal, Canada. His current research interests include fog and cloud computing, Artificial intelligence, Machine learning, Federated Learning, and Cyber Security.
\endbio

\vskip60pt
\bio{PublicationsNewPic2021}
Azzam Mourad is currently a Visiting Professor at Khalifa University, a Professor of Computer Science and Founding Director of the Artificial Intelligence and Cyber Systems Research Center at the Lebanese American University, and an Affiliate Professor at the Software Engineering and IT Department, Ecole de Technologie Superieure (ETS), Montreal, Canada. He was a Visiting Professor at New York University Abu Dhabi. His research interests include Cyber Security, Federated Machine Learning, Network and Service Optimization and Management targeting IoT and IoV, Cloud/Fog/Edge Computing, and Vehicular and Mobile Networks. He has served/serves as an associate editor for IEEE Transactions on Services Computing, IEEE Transactions on Network and Service Management, IEEE Network, IEEE Open Journal of the Communications Society, IET Quantum Communication, and IEEE Communications Letters, the General Chair of IWCMC2020-2022, the General Co-Chair of WiMob2016, and the Track Chair, a TPC member, and a reviewer for several prestigious journals and conferences. He is an IEEE senior member.
\endbio
\vskip3pt

\bio{hadi}
Hadi Otrok (senior member, IEEE) received his Ph.D. in ECE from Concordia University, Montreal, QC, Canada, in 2008.He holds a Full Professor position in the Department of Computer Science at Khalifa University, Abu Dhabi, UAE. He is also an Affiliate Associate Professor in the Concordia Institute for Information Systems Engineering at Concordia University, and an Affiliate Associate Professor in the Electrical Department at Ecole de Technologie Superieure (ETS), Montreal, Canada. His research interests include the domain of blockchain, reinforcement learning, federated learning, crowd sensing and sourcing, ad hoc networks, and cloud security. He co-chaired several committees at various IEEE conferences. He is also an Associate Editor at IEEE Transactions on Network and Service Management (TNSM), IEEE Transactions on Service Computing, and Ad-hoc networks (Elsevier). He also served in the editorial board of IEEE Networks and IEEE Communication Letters.

\endbio
\vskip30pt

\bio{jamal}
Jamal Bentahar received the Ph.D. degree in computer science and software engineering from Laval University, Canada, in 2005. He is a Professor with Concordia Institute for Information Systems Engineering, Concordia University, Canada and visiting professor at Khalifa University, UAE. From 2005 to 2006, he was a Postdoctoral Fellow with Laval University, and then NSERC Postdoctoral Fellow at Simon Fraser University, Canada. He was an NSERC Co-Chair for Discovery Grants for Computer Science (2016–2018). He is an associate editor of IEEE Transactions of Services Computing. His research interests include computational logics, model checking, reinforcement and deep learning, multi-agent systems, and services computing.
\endbio
\vskip30pt
\clearpage
\bio{mohsen}
Mohsen Guizani (Fellow, IEEE) received the BS (with distinction), MS and PhD degrees in Electrical and Computer engineering from Syracuse University, Syracuse, NY, USA in 1985, 1987 and 1990, respectively. He is currently a Professor of Machine Learning and the Associate Provost at Mohamed Bin Zayed University of Artificial Intelligence (MBZUAI), Abu Dhabi, UAE. Previously, he worked in different institutions in the USA. His research interests include applied machine learning and artificial intelligence, Internet of Things (IoT), intelligent autonomous systems, smart city, and cybersecurity. He was elevated to the IEEE Fellow in 2009 and was listed as a Clarivate Analytics Highly Cited Researcher in Computer Science in 2019, 2020 and 2021. Dr. Guizani has won several research awards including the “2015 IEEE Communications Society Best Survey Paper Award”, the Best ComSoc Journal Paper Award in 2021 as well five Best Paper Awards from ICC and Globecom Conferences. He is the author of ten books and more than 800 publications. He is also the recipient of the 2017 IEEE Communications Society Wireless Technical Committee (WTC) Recognition Award, the 2018 AdHoc Technical Committee Recognition Award, and the 2019 IEEE Communications and Information Security Technical Recognition (CISTC) Award. He served as the Editor-in-Chief of IEEE Network and is currently serving on the Editorial Boards of many IEEE Transactions and Magazines. He was the Chair of the IEEE Communications Society Wireless Technical Committee and the Chair of the TAOS Technical Committee. He served as the IEEE Computer Society Distinguished Speaker and is currently the IEEE ComSoc Distinguished Lecturer. 
\endbio

\end{document}